\documentstyle[eqsecnum,prd,aps]{revtex}
\input epsf

\def\prl{Phys.\ Rev.\ Lett.}
\def\pr{Phys.\ Rev.}

\def\nat{Nature}
\def\pl{Phys.\ Lett.}
\def\np{Nucl.\ Phys.}

\def\apj{Ap.\ J.}
\def\apjl{Ap.\ J.\ Lett.}

\def\mn{M.$\,$N.$\,$R.$\,$A.$\,$S.}

\def\jetp{Sov.\ Phys.\ JETP}

\def\camup{Cambridge University Press}

\begin{document}
\draft
\title{Galactic Magnetic Fields from Superconducting Strings}
\author{C. J. A. P. Martins\thanks{Also at C. A. U. P.,
Rua do Campo Alegre 823, 4150 Porto, Portugal.
Electronic address: C.J.A.P.Martins\,@\,damtp.cam.ac.uk}
and
E. P. S. Shellard\thanks{Electronic address:
E.P.S.Shellard\,@\,damtp.cam.ac.uk\vskip0pt Submitted to
Phys.\ Rev.\ {\bf D}.}}
\address{Department of Applied Mathematics and Theoretical Physics\\
University of Cambridge\\
Silver Street, Cambridge CB3 9EW, U.K.}
\date{23 May 1997}
\maketitle

\begin{abstract}
We use a simple analytic model for the evolution of currents in superconducting
strings to estimate the strength of the `seed' magnetic fields generated
by these strings. This model is an extension of the evolution model of
Martins and Shellard depending on a parameter $f$ which characterizes the
importance of equilibration process in the evolution of the currents.
For GUT-scale strings, we find that a viable seed magnetic field for the
galactic dynamo can be generated if equilibration is weak.
On the other hand, electroweak-scale strings originate magnetic fields
that are smaller than required.
\end{abstract}

\section{Introduction}
\label{b-int}
All observations show that our galaxy, together with a good number of other
spirals (and galaxy clusters in the inter-cluster medium),
possess `regular' magnetic fields with magnitude $B\sim10^{-6}\,G$, on scales
of several kiloparsecs \cite{zel,ruz,revs} (in addition, there is a
small-scale random component in our galaxy with the same magnitude and a
coherence length of about $100\,Mpc$). No magnetic fields
have been observed on larger scales, current observational bounds (obtained
from the analysis of remote radio galaxies and quasars) being about
$B<10^{-9}\,G$ \cite{val}---but it should be said that, since
one needs to separate between source, Galaxy and intergalactic contributions,
these are quite difficult observations. Even though these magnatic fields
are fairly small, it is of course possible to find localized objects with
much larger magnetic fields---for example, X-ray sources near neutron stars
can have  $B\sim10^{13}\,G$.

These galactic fields are associated with the interstellar gas. Stellar
magnetic fields are known to be extremely small between stars, and in
any case they could not
explain the observed large-scale structure. Even though the magnetic fields
do not play any significant part in the equilibrium and dynamics of the
galaxy, they do have a significant role in the propagation of cosmic rays,
gasdynamical processes---notably star formation---and in the mechanism
by which cosmic dust is oriented. In particular, star formation is not possible
without a magnetic field---its role being that of transporting angular momentum
outwards so that the collapse of the proto-stellar cloud can continue.

The large coherence scales of these magnetic fields (several kiloparsecs) means
that it is difficult to find mechanisms capable of createing them (for example,
thermal, chemical or other `battery' effects are inadequate).
Thus, even though a large number of possibilities have been considered in the
past---including vorticity \cite{harri}, inflationary
models\cite{ratra,quash} and
cosmological phase transitions \cite{hogan}---none seems to be
particularly compelling.

In this paper we consider the possibility of the galactic magnetic
fields \cite{zel,ruz,revs} being generated by superconducting \cite{witten,vs}
cosmic strings. Our discussion is based on the quantitative evolution model of
Martins and Shellard \cite{ms,ms1}, together with a simple `toy model' for the
evolution of the superconducting currents \cite{mscur,msvor}.
We consider both electroweak- and GUT-scale cosmic strings. While earlier
estimates indicated that superconducting GUT strings were observationally
ruled out, since they led to unacceptebly large densities of springs and
vortons \cite{vor}, it has been shown---by Peter \cite{ppeters} for the former,
by Martins and Shellard \cite{msvor} for the later---that neither of these
form in general.

The plan of this paper is as follows. In section \ref{b-str} we review some
basic notions about astrophysical and cosmological magnetic fields. Following
this we briefly review our evolution model (first discussed in
\cite{mscur}, see also \cite{msvor}) in section \ref{b-evc} and
analise its solutions. In section \ref{b-bbb} we determine the relevant `seed'
magnetic fields and compare our results with existing bounds; finally
(section \ref{b-con}), we discuss the relevance of our results.

\section{Astrophysical Magnetic Fields}
\label{b-str}
It has been shown \cite{zel,ruz,revs} that in order to explain the observed
galactic magnetic fields $B_o\sim10^{-6}\,G$ one needs a seed
field $B_s\ge10^{-19}\,G$
on the comoving scale of a protogalaxy (about $100\,kpc$)---such field can
then be amplified, by a dynamo mechanism, to the observationally required
value. Since the gravitational collapse of the protogalaxies enhances any
frozen-in magnetic field, this seed field corresponds to an rms field
\begin{equation}
B_g\ge10^{-22}\,G\, \label{bgal}
\end{equation}
at the epoch when galactic scales $d_g\sim1\,Mpc$ fall inside the
horizon (that is, at a time $t_g\sim2.5\times10^{-3}\,t_{eq}$). On the other
hand, one would need $B_g\sim 10^{-10}G$ to create significant magnetic fields
through adiabatic compression alone.

We should recall that there are also upper bounds
on cosmologically interesting magnetic fields. Firstly, primordial
nucleosynthesis is sensitive to magnetic fields, which change the expansion
rate of the universe an consequently the rates of the reactions
that produce the light elements.
This gives rise to a bound \cite{nucl}
\begin{equation}
B_{nuc}\le10^9\,G\, ,\label{bnuc}
\end{equation}
or equivalently
\begin{equation}
\left(\frac{\rho_B}{\rho_\gamma}\right)_{nuc}\le0.28\, \label{bnucden}
\end{equation}
at the nucleosynthesis epoch.
Secondly, there are bounds on the strength of a uniform `primordial' magnetic
field; as we pointed out in the previous section, analysis of radio
galaxies and quasars yields the constraint \cite{val}
\begin{equation}
B_0\le10^{-9}\,G\, .\label{btod}
\end{equation}
A more recent analysis of the consequences of primordial magnetic fields for
the cosmic microwave background \cite{bfs} produces a comparable bound
\begin{equation}
B_0\le6.8\times10^{-9}\Omega_0^{1/2}h\,G\, ;\label{btodcmb}
\end{equation}
in terms of densities, these can be written
\begin{equation}
\left(\frac{\rho_B}{\rho_\gamma}\right)_0\le10^{-7}\, .\label{btodden}
\end{equation}

It should be pointed out that an {\em ab initio} uniform magnetic field does
not violate homogeneity, but it does make the cosmological expansion
anisotropic. It is thus much more appealing to assume that the required
magnetic seed fields were generated by some dynamical process. The currently
favoured paradigm is  dynamo theory\cite{zel,ruz}, which
develops on Larmor's suggestion\cite{larm} that
it is possible to excite magnetic fields by the motion of a conductive
fluid in a gas, and allows energy associated with the differential
rotation of spiral galaxies to
be converted into magnetic field energy. In this model the state of the
magnetic field today is almost independent of initial conditions:
a dynamo process results in equipartition of energy between the plasma
kinetic and magnetic energies on scales up to the coherence length of the
field.

The first proposal for the origin of the required seed field was originally
due to Harrison\cite{harri}, and subsequently developed by Mishustin and
Ruzmaiakin \cite{mishu}. This claims that the relative motion of protons and
electrons induced by vorticity present before the epoch of decoupling
(the electrons in vortices being more strongly coupled to the background
radiation than the protons) produces
primeval currents and hence magnetic fields. Obviously, this requires a source
of vorticity.

Later Vachaspati and Vilenkin\cite{vacha} have suggested that strings with
small-scale structure are such a source. They pointed out that, since the
matter flow in baryonic wakes is turbulent,
velocity gradients will be induced in the flow by the small-scale wiggles,
which produces the required vorticity.
Avelino and Shellard\cite{ppa} have also shown that dynamical friction between
cosmic strings and matter provides a further source of vorticity.

It is also possible to generate large-scale magnetic fields at the end
of an inflationary epoch\cite{ratra,quash}. However, these models generally
need to invoke rather speculative changes to the nature of the electromagnetic
interactions during the inflationary epoch, whose only motivation seems
to be the generation of such magnetic fields.

Still, none of the above (or other) possibilities provides a compelling
mechanism for the generation of a magnetic field with strength $B_g$
on galactic scales $d_g$ today. We should also point out that there is evidence
that magnetic fields were present in moderately young galaxies (at redshifts
$z\sim1$--$2$) \cite{revs}. This is a challenge to dynamo theory, in that at
least the simplest galactic dynamo models cannot generate micro-Gauss strength
magnetic fields at such early epochs \cite{wolfe}.

\section{Evolution of the String Network}
\label{b-evc}
Due to the strings' statistical nature, analytic evolution methods must be
`thermodynamic', that is one must describe the network by a small number
of macroscopic (or `averaged') quantities whose evolution equations
are derived from the microscopic string equations of motion. The first such
model providing a quantitative picture of the
complete evolution of a string network (and the corresponding loop
population) has been developed by Martins and
Shellard (see \cite{ms,ms1} for a detailed analysis of the model), and has
two such quantities, the long-string correlation length
$\rho_{\infty}\equiv\mu/L^2$ ($\mu$ being the string mass per unit length)
and the string RMS velocity,
$v^2\equiv\langle{\dot{\bf x}}^2\rangle$.
It also includes two `phenomenological' parameters, a `loop chopping
efficiency' $0<{\tilde c}<1/2$ and a `small-scale structure
parameter' $0<k<1$. These are sufficient to quantitatively
describe the large-scale properties of a cosmic string network.

More recently, this has been extended with a `toy model' for the evolution
of the superconducting currents (see \cite{mscur,msvor}). Assuming that there
is a `superconducting correlation length', denoted $\xi$, which measures the scale over which
one has coherent current and charge densities on the strings, we can define
$N$ to be the number of uncorrelated current regions
(in the long-string network) in a co-moving volume $V$. It is then fairly
straightforward to see how the dynamics of the string network affects $N$ and
obtain an evolution equation for it. The only non-trivial issue is that of
the dynamics of the currents themselves. There is evidence that some kind of
`equilibration' process acts between neighbouring current regions,
counteracting the creation of new regions by inter-commutings and helping
their removal by loops. Notably, the simulations of
Laguna and Matzner \cite{laguna} show that as the result of inter-commutings
charges pile up at current discontinuities and move with the kinks, but their
strength decreases with time. Also, Austin, Copeland
and Kibble have shown \cite{ack} that in an expanding universe correlations
between left- and right-moving modes develop due both to stretching and
inter-commuting (particularly when loops form). We model this term by
assuming that after each Hubble time, a fraction $f$ of the $N$ regions
existing at its start will have equilibrated with one of its neighbours,
\begin{equation}
\left(\frac{dN}{dt}\right)_{dynamics}=-fHN\, ;
\label{eqlterm}
\end{equation}
note that new regions are obviously created by inter-commuting during the
Hubble time in question, so that $f$ can be larger than unity.
Alternatively we can say that for a given $f$, the number of regions that were
present in a
given volume at a time $t$ will have disappeared due to equilibration at a
time $t+(fH)^{-1}$.

We therefore obtain the following evolution equation for $N$
\begin{equation}
\frac{dN}{dt}=G\left(\frac{\ell}{\xi}\right)
\frac{v_\infty}{\alpha}\frac{V}{L^4}-fHN\, ,
\label{generalnn}
\end{equation}
where the `correction factor' $G$ has the form (see \cite{msvor} for a
complete discussion)
\begin{equation}
G\left(\frac{\ell}{\xi}\right)=\left\{ \begin{array}{ll}
2-{\tilde c}\left(\frac{\ell}{\xi}+2\right) \, ,&
\mbox{$\frac{\ell}{\xi}>1$} \\
2(1-2{\tilde c})\alpha+(2-3{\tilde c}-2\alpha+4{\tilde c}\alpha)
\frac{\ell}{\xi} \, ,& \mbox{$\frac{\ell}{\xi}\le1$} \end{array} \right. \, ;
\label{gansg}
\end{equation}
loops are assumed to form with a size $\ell(t)=\alpha(t)L(t)$, where
$\alpha\sim1$ while the string network is is the friction-dominated epoch
and $\alpha=\alpha_{sc}\ll1$ once it has reached the linear scaling regime
(see \cite{ms1}).

For what follows it is more convenient to introduce $N_L$, defined to be
the number of uncorrelated current regions per long-string correlation length,
\begin{equation}
N_L\equiv\frac{L}{\xi}\, ;\label{defnl}
\end{equation}
in terms of $N_L$, (\ref{generalnn}) has the form
\begin{equation}
\frac{dN_L}{dt}=(3v_\infty^2-f)HN_L+\frac{3}{2}
\frac{v_\infty^2}{\ell_{\rm f}}N_L+\left(\frac{1}{\alpha}G(\alpha N_L)+
\frac{3}{2}{\tilde c}N_L\right)\frac{v_\infty}{L}\, ,
\label{generalnl}
\end{equation}
where $\ell_{\rm f}$ is the friction lengthscale due to particle scattering
off strings. This has been shown \cite{mscur} to be the dominant friction
mechanism, except possibly if there are background magnetic fields (in
which case plasma friction effects would be more important). Such possibility
will not be considered in this paper, since we are interested in the magnetic
fields generated by the strings themselves. Note that to obtain this
equation one needs to use the evolution equation for the long-string
correlation length $L$, and that one can equivalently define $G$ as
\begin{equation}
G\left(\alpha N_L\right)=\left\{ \begin{array}{ll}
2-{\tilde c}\left(\alpha N_L+2\right) \, ,&
\mbox{$\alpha N_L>1$} \\
2(1-2{\tilde c})\alpha+(2-3{\tilde c}-2\alpha+4{\tilde c}\alpha)
\alpha N_L \, ,& \mbox{$\alpha N_L\le1$} \end{array} \right. \, .
\label{newgans}
\end{equation}

Now the question is, of course, what is $f$.
From a more intuitive point of view, an equivalent question is the following:
given a particular piece of string with a given current, is it more likely to
disappear from the network by this equilibration mechanism or by being
incorporated in a loop? Even though a precise answer can probably only be
given by means of a numerical simulation, some physical arguments can be
used to constrain it \cite{mscur,msvor}. In the present paper, however, we
will postpone this interesting discussion and treat $f$ as a free parameter.
We simply point out that, according to our previous results \cite{mscur,msvor},
if equilibration is inexistent or ineffective, then $N_L$ grows without bound
at late times, whereas if equilibration is effective $N_L$ eventually becomes
a constant (which corresponds to linear scaling of $\xi$).

We should also say at this stage that once the network leaves the
friction-dominated regime and strings become relativistic other mechanisms
(notably radiation) can cause charge losses in the long strings (as well as
in loops). Thus we do not expect our toy model to provide quantitatively
correct answers, but we do expect it to provide reliable order-of-magnitude
estimates.

\section{Magnetic fields}
\label{b-bbb}
We expect the seed magnetic fields from cosmic strings to be coherent on the
scale at which loops are being formed, that is $\ell(t)=\alpha(t) L(t)$.
Hence if $N_L$ is the number of uncorrelated regions per long-string
correlation length and $L(t)=\gamma(t) t$ we find
\begin{equation}
B_g=\frac{2\pi e}{c^2t_g^2\alpha^{3/2}}\,\frac{N_L^{1/2}}{\gamma^2}\, ,
\label{fields}
\end{equation}
and all we have to do is evaluate $N_L$ and $\gamma$ using our analytic model,
while checking that at the nucleosynthesis epoch the corresponding magnetic
fields are consistent with existing bounds---which is isdeed the case.

\begin{figure}
\vbox{\centerline{
\epsfxsize=1.0\hsize\epsfbox{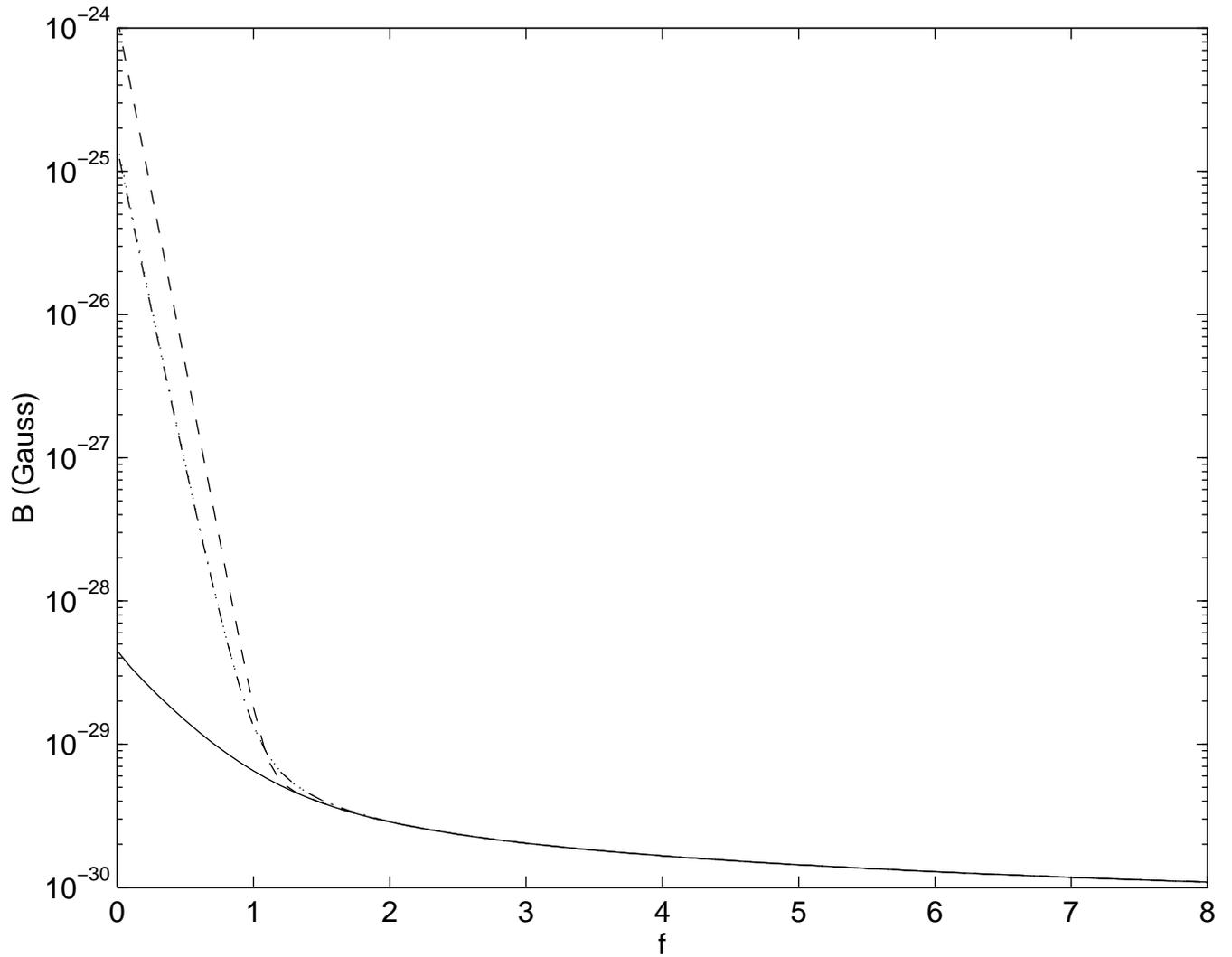}}
\vskip.4in}
\caption{The magnitude of coherent magnetic fields, at the epoch when galaxy
scales fall inside the horizon, as a function of $f$ for electroweak-scale
superconducting string networks. The different lines correspond to initial
conditions typical of string-forming and superconducting phase transitions
that are respectively of 1st \protect\& 1st (solid), 1st \protect\& 2nd
(dashed), 2nd \protect\& 1st (dash-dotted) and 2nd \protect\& 2nd (dotted)
order.}
\label{bds_e}
\end{figure}

\begin{figure}
\vbox{\centerline{
\epsfxsize=1.0\hsize\epsfbox{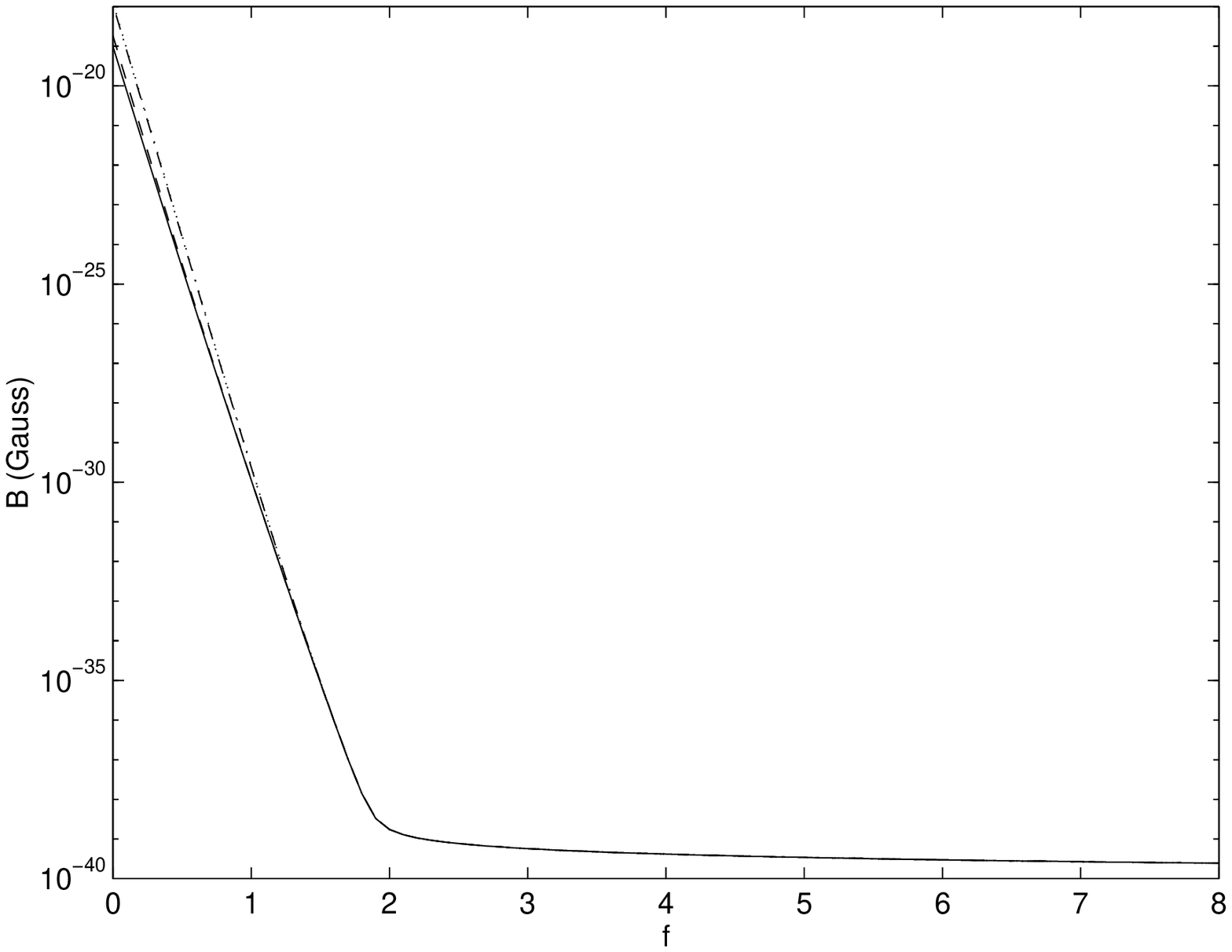}}
\vskip.4in}
\caption{The magnitude of coherent magnetic fields, at the epoch when galaxy
scales fall inside the horizon, as a function of $f$ for GUT-scale
superconducting string networks. The different lines correspond to initial
conditions typical of string-forming and superconducting phase transitions
that are respectively of 1st \protect\& 1st (solid), 1st \protect\& 2nd
(dashed), 2nd \protect\& 1st (dash-dotted) and 2nd \protect\& 2nd (dotted)
order.}
\label{bds_g}
\end{figure}

In figures \ref{bds_e} and \ref{bds_g} we plot the expected coherent seed fields
at the epoch $t_g$, for $0\le f\le8$---as can be seen, for large enough $f$ the
result is almost independent of it. It can be seen that if equilibration
is ineffective
electroweak strings just fall short of producing the required seed fields,
$B_{seed}\sim10^{-22}\,Gauss$, but GUT-scale strings can in the same
circumstances produce such fields---all we require is an ineffective
equilibration mechanism, $f\le0.5$.

Note that there is almost no dependence
on initial conditions in the GUT case, but such dependence persists for the
electroweak string network if equilibration is weak. this is because
the GUT-scale string network is in the linear scaling regime at $t_g$,
while the electroweak network is in the Kibble regime (see \cite{ms1} for
a detailed description of these regimes).
In the GUT case, $N_L$ is constant (that is, $\xi$ is caling linearly) provided
$f\ge1.88$ (see \cite {mscur,msvor}), whereas if $f=0$ $\xi$ is constant
and $N_L$ is growing linearly. This explains the large differences between the
magnetic fields at high and low $f$ in the GUT case compared to the electroweak
one. On the other hand, $t_g\sim2.5\times 10^{47}\,t_{GUT}$ and
$t_g\sim2.5\times 10^{19}\,t_{EW}$ in the GUT and electroweak cases
respectively. Thus, despite evolving for a much shorter time, electroweak
strings are friction dominated much longer than GUT-ones, and so if
equilibration is effective they can build-up much larger currents. This is the
reason why in this case electroweak strings generate much larger
magnetic fields.

\section{Conclusions}
\label{b-con}
In this paper we have used the quantitative string evolution model of
Martins and Shellard \cite{ms,ms1}, together with a simple toy model for the
evolution of currents on the strings \cite{mscur,msvor} to study the possibilty
of using superconducting strings to provide the `seed' galactic
magnetic fields.

We have shown that GUT-scale superconducting strings can provide
the required fields for the galactic dynamo mechanism provided that current
equilibration mechanism are ineffective, while similar fields from
electroweak strings are too weak.

Clearly, the outstanding issue, in this and other cosmological scenarios
involving superconducting cosmic strings, is that of the importance of charge
and current equilibration mechanisms on the strings, and a more detailed
study of it is therefore required.

\acknowledgments
C.M.\ is funded by JNICT
(Portugal) under `Programa PRAXIS XXI' (grant no.
PRAXIS XXI/BD/3321/94). E.P.S.\ is funded by PPARC and
we both acknowledge the support of PPARC and the
EPSRC, in particular the Cambridge Relativity rolling
grant (GR/H71550) and a Computational Science
Initiative grant (GR/H67652).

\vfill\eject

\vfill\eject

\end{document}